\documentclass[aps,13pt,eqsecnum, preprint,nofootinbib,superscriptaddress]{revtex4}
\usepackage{amssymb,amsmath,amsthm,graphicx,amscd}
\usepackage{enumerate,color,changepage,ulem,bm}
\usepackage[hidelinks]{hyperref}

\DeclareMathSymbol{\shortminus}{\mathbin}{AMSa}{"39}

\newcommand{\eqn}[1]{(\ref{#1})}

\begin{document}


\title{NonMarkovianity in  Cosmology: Memories kept in a Quantum Field}

\author{Jen-Tsung Hsiang}
\email{cosmology@gmail.com}
\affiliation{Center for High Energy and High Field Physics, National Central University, Taoyuan 320317, Taiwan, ROC}
\author{Bei-Lok Hu}
\email{blhu@umd.edu}
\affiliation{Maryland Center for Fundamental Physics and Joint Quantum Institute,  University of Maryland, College Park, Maryland 20742, USA}

\begin{abstract}
In this work we ask how an  Unruh-DeWitt (UD) detector with harmonic oscillator internal degrees of freedom $Q$ measuring an evolving quantum matter field $\Phi (\bm{x}, t)$ in an expanding universe with scale factor $a(t)$ responds. We investigate the detector's response which contains non-Markovian information about the quantum field squeezed by the dynamical spacetime. 
The challenge is in the memory effects accumulated over the evolutionary history. 
We first consider a detector $W$, the `\textsl{Witness}', which co-existed and evolved with the quantum field from the beginning. 
We derive a nonMarkovian quantum Langevin equation for the detector's $Q$ by integrating over the squeezed quantum field. The solution of this integro-differential equation would answer our question, in principle, but very challenging, in practice.  Striking a compromise, we then ask, to what extent can a detector $D$ introduced at late times, called the `\textsl{Detective}', decipher past memories. This situation corresponds to many cosmological experiments today probing specific stages in the past, such as COBE targeting activities at the surface of last scattering.  Somewhat surprisingly we show that it is possible to retrieve to some degree certain global physical quantities, such as the resultant squeezing, particles created, quantum coherence and correlations. The reason is because the quantum field has all the fine-grained information from the beginning in how it was driven by the cosmic dynamics $a(t)$. How long the details of past history can persist in the quantum field depends on the  memory time. The fact that a squeezed field cannot  come to complete equilibrium under continuous driving,  as in an evolving spacetime,  actually helps to retain the memory. We discuss interesting features and potentials of this `\textit{archaeological}' perspective toward cosmological issues. 
\end{abstract}


\maketitle

\hypersetup{linktoc=all}
\setcounter{tocdepth}{2}
\tableofcontents
\clearpage
\baselineskip=18pt
\numberwithin{equation}{section}
\allowdisplaybreaks

\section{Introduction}

Ever since a moving detector was introduced by    Unruh \cite{Unr76} to depict analog Hawking effect \cite{Haw74,HawHar76} it has become a useful means to extract information about a quantum field, as is done in many work in relativistic quantum information \cite{RQI}. For cosmology a detector is often used to illustrate the Gibbons-Hawking effect \cite{GibHaw} an observer may find in de Sitter space. Here we treat a more general class of problems allowing for an arbitrary functional dependence of the cosmic scale factor $a(t)$ between two constant end states and not restricting our attention to thermal radiation.  We ask how an  Unruh-DeWitt (UD) detector \cite{Unr76,DeW79} with harmonic oscillator internal degrees of freedom $Q$ measuring an evolving quantum matter field $\Phi$ driven by the expansion of the universe would respond. The dynamical response of the detector contains information about the quantum field which is squeezed over the history of the universe. It is well-known that such nested dependence on $\Phi (\bm{x}, t)$ and then on $a(t)$ gives rise to nonMarkovian dynamics of $Q$. 
The question we pose here is, to what extent can a detector introduced at a later time in the history of the universe extract stored information about the matter field and from it the dynamics of the cosmos at an early time, as well as the information of the parametric process of the field.  The challenge is in the memory effects accumulated in the evolutionary history. 

The problem we set forth to investigate consists of three components: 
a) cosmological expansion results in the squeezing of the quantum field and manifests as particle creation, b) the nonequilibrium  dynamics of an UD detector in a squeezed quantum field, c) the dynamical response of the detector  to the quantum matter field which has a nonMarkovian history. The first component is well known from work since the 70s. The second component we shall describe below. The third component is where the major challenge rests. The background themes are well-researched with many monographs dedicated to them: \textit{squeezed states} in quantum optics \cite{Walls,LouKni,ManWol}, \textit{open quantum systems} \cite{qos}, \textit{nonequilibrium quantum field dynamics} \cite{CalHu08,NEqFT}, and \textit{cosmological particle creation} in quantum field theory in curved spacetime \cite{BirDav,ParToms,HuVer20}. We shall give a brief description of these three major components,  extending our earlier work and referring to the  current literature. We then focus on the nonequilibrium dynamics of a UD detector interacting with a quantum field under time-dependent squeezing, here, due to the expansion of the universe, paying special attention to the nonMarkovian behavior.

\subsection{Early universe quantum processes}

As examples of cosmological problems involving squeezed quantum fields we mention three processes in the early universe. Then we introduce an UD detector, describe its nonequilibrium dynamics and discuss what it sees in various situations. The theoretical framework useful for our purpose is known as `squeezed open quantum systems' \cite{KMH97}.\\	

\noindent 1) {\it Cosmological particle creation as squeezing of the quantum field}  


The amplitude of a wave mode can be parametrically amplified by a time-dependent drive.  Same thing happens to vacuum fluctuations in a quantum field, resulting in the creation of particle pairs \cite{Schwinger}. The expansion of the universe acts like a drive, parametrically amplifying the quantum noise leading to cosmological particle creation \cite{Par69,Zel70}. The vacuum is `squeezed' in the evolutionary history and particles are produced \cite{GriSid} -- this is spontaneous creation.  If there were particles present in an initial state they get amplified also with the same amplification factor -- this is stimulated production. A summary description of cosmological particle creation in terms of squeezing can be found in \cite{HKM94}. By applying the  well established knowledge base about squeezing in quantum optics one can observe or design experiments simulating quantum processes in the early universe (e.g., \cite{CalHu04}) and in black holes (e.g., \cite{Garay}). This is the spirit of analog gravity \cite{analogG} invoking the similarity of the key physical processes and  the  commonality of the underlying issues.  \\

\noindent 2) \textit{Initial state of inflation or phase of the universe}  

Making use of the properties of squeezing on quantum states Agullo and Parker \cite{AguPar} considered an initial mixed state with 
nonzero numbers of scalar particles present and calculated the induced stimulated production of density perturbations by the inflationary expansion of the universe from early times.  They found that  the effect of these initial perturbations is not diluted by inflation and can  significantly enhance non-Gaussianities in the  squeezed limit. Observations of these non-Gaussianities, they claim,  can  provide valuable information about the initial state of the inflationary universe.  In a similar spirit the authors of \cite{GMM14} claim that ``the existence (or not) of a quantum bounce leaves a trace in the background quantum noise that is not damped and would be non-negligible even nowadays." 
\\ 


\noindent 3) \textit{Entropy generation} 

Entropy of quantum fields is a fundamental yet somewhat tricky issue. Early conceptual inquiries \cite{HuPav,HuKan} made explicit the underlying conditions which allow for the entropy of free quantum fields to be related to particle creation numbers (for boson fields). Namely, by adopting a Fock space representation and making statements only in terms of number operators, one implicitly ignores all quantum phase information which determines the coherence. (See \cite{KME} for the interplay of both quantities). From this it is easy to understand why the entropy associated with particle production is proportional to the degree of squeezing by the expansion of the universe \cite{GasGio93,Prokopec93,KMH97}. The relation between entanglement and entropy in particle creation is explicated in \cite{LCH10}.  Entropy associated with gravitational perturbations in the inflationary universe is explored in \cite{BMP92,GGV93,KPS00,AMM05,CamPar,KPS12} and by many authors (see  \cite{Bran20} for an extensive literature and the latest developments in terms entanglement entropy). It was suggested in \cite{GGV93} 
that the entropy produced by the squeezing of the universe is independent of its initial state. One can then ask, does this mean by measuring this entropy  today one can find the accumulated degree of squeezing, and from it deduce the duration of the last inflationary expansion {regardless of the initial conditions }?  

We have suggested some samples of measurable physical quantities of a quantum field in a dynamical setting. Quantum decoherence, correlation and entanglement of gravitational perturbations in cosmology are related important subjects  (see, e.g., \cite{LCH10,MarMenCosEnt,Bran20}  and references therein).  We shall discuss these from a detector response in squeezed quantum field perspective in  later communications \cite{HHCosDec,HHCosEnt}.    Introducing a detector to measure the quantum field driven by cosmological expansion can extend our investigative capabilities, as the detector response function registers both the matter field and the cosmic activities. For that, we need some knowledge of the nonequilibrium dynamics of the quantum field and the detector.

\subsection{Nonequilibrium quantum dynamics}

There are three players in the problem we are tackling: The expanding universe acting as a drive, the quantum field which responds to it in processes such as those listed above, and the UD detector, which we assume to be a harmonic oscillator. 
We want to find out from the nonMarkovian response of the detector and the nonequilibrium dynamics of the quantum field how much we can unlock the past history of the universe.

\paragraph{`Witness' versus `Detective'}

Here we make a distinction between two kinds of observers represented by an UD detector which is a simple harmonic oscillator with time-dependent natural frequency: one which has co-existed and evolved with the matter which we shall call the `witness' $W$. The other is introduced in a much later stage of cosmic evolution, which we call the `detective' $D$. It is what we design in all the cosmological experiments today to find out what could have happened earlier. For example, the detectors in COBE, WMAP and PLANCK are aimed at observing the radiation emitted from the surface of last scattering.  So what is a witness? Witnesses are physical objects present from the inception which followed the evolution and are present today (or at the moment of detection). Examples are the heavy elements on earth which contain the histories of their journeys to here since their formation in the interiors of the neutron stars. In cosmology, they could be the light elements from nucleosynthesis, or baryons from baryogenesis or particles produced from vacuum fluctuations described earlier. They are quantities of relevance to what we can observe today which evolved alongside the expansion of the universe from the earliest moments. It could be the Planck time if one is interested in quantum gravity effects, or the GUT time if one believes that is the most important inflationary transition which can affect the main features of today's universe.  Density contrasts and primordial gravitational waves, related to the scalar and tensor sectors of gravitational perturbations, are two important examples. Let us see this more explicitly.

\paragraph{Gravitational Scalar and Tensor Perturbations}

For  density contrasts and gravitational perturbations: from the Einstein equations (e.g. Eq. (28.47) of \cite{ThoBla}) 
for the gravitational potentials $\Phi$, $\Psi$ (in Eq. (28.44) \textit{ibid}) the equations governing their normal modes, with suitable choice of gauges, obey equations of motion in a harmonic oscillator form with time-dependent natural frequency (E.g. Eq. (28.50) \textit{ibid}, with time measured by $\ln a (t)$, the scale factor, further simplified to Eq. (28.51) \textit{ibid}, with $\Phi = \Psi$ for a radiation dominated background). 
For gravitational waves it is long known (e.g., \cite{ForPar77}) that the two polarizations $h_+$, $h_{\times}$ each obey  an equation like that of a massless minimally-coupled scalar field. The equation of motion of each normal mode in the Heisenberg picture is given by, e.g., Eq. (4.12) in \cite{FRV12} in the form of an equation for a harmonic oscillator with time-dependent frequency\footnote{These equations all have a first order time derivative term but by a change of variables can be cast to a form with only the second order time derivative term and a zero-order term with time-dependent frequency.}, with solution given by (4.14) in \cite{FRV12}. See\footnote{As an explanatory note in the differences between `nonMarkovian' or memory effects from that of `secular' effects it is enough to adopt the description of \cite{KapBur}:
``Secular growth is the phenomenon where the coeffients $c_n(t)$ of a perturbative evaluation of some observable in powers of some small coupling $|g|\ll 1$, are time-dependent and grow without bound at late times." Our treatment here is nonperturbative and we do not need to assume weak coupling. NonMarkovian refers to any process with some memory, as signified by a nonlocal kernel in the equations of motion. Integrating an integro-differential equation forward one time step requires the knowledge not only of the immediate past (Markovian limit), but also data in the remote past, to varying degrees of nonMarkovianity.} also ~\cite{Wu11,HF17}.   

In the first part of this paper we shall treat the nonequilibrium dynamics of a harmonic oscillator with time-dependent frequency, i.e., an UD detector  regarded as the Witness, representing the physical quantities of interest as exemplified above. The nonMarkovian quantum Langevin equation is obtained by  integrating over the quantum field which is squeezed by the expansion of the universe. Notice that the quantum field keeps a record of how it was squeezed, and the nonequilibrium dynamics of the $W$ detector contains the coarse-grained (`integrated over') information of the field. Thus the answer to the question we posed -- is it possible to retrieve some information about the early universe -- is answered by solving this integral-differential equation. We present the equation, with some discussion of the nonMarkovian feature, but leave the complete solutions to some later more able hands.


\subsection{Detector response to nonMarkovian dynamics} 

In \cite{Unr76} the detector \textit{response function} refers to the field's Wightman function while the detector's  \textit{selectivity} refers to its internal energy change (such as the excitation from a ground state to an excited state). The theoretical framework employed there is time-dependent perturbation theory (TDPT). Along this line there is a huge literature on this subject. See, e.g., \cite{Higuchi,GarProRes,GarProUni,LouSat} and references therein. 

\paragraph{Perturbative methods inadequate for nonMarkovian processes}
Most prior work  on UD detectors are based on TDPT, which is valid only for very weak couplings between the detector and the field and a short time after the detector-field coupling is switched on, and since  the perturbative results become inaccurate quickly,  any backreaction calculation of the field on the detector based on perturbation theory becomes unreliable  just as quickly,  incapacitating it to treat nonMarkovian effects\footnote{There were suggestions to go nonperturbative, e.g., \cite{BMMM13,BLF13}.  The research program of  one of us with collaborators on detector-field interactions since the 90s has emphasis in exact treatments,   from the derivation of an exact nonMarkovian master equation for quantum Brownian motion  \cite{HPZ,HM94},  two field interactions  in de Sitter universe \cite{HPZdec,Banff}  for the study of quantum decoherence,  the uncertainty relation at finite temperatures \cite{HZ93},  squeezing and entropy  generation \cite{HKM94,KMH97}, field-induced mutual nonMarkovianity in multiple-detectors \cite{RHA,CPR}  to  backreaction \cite{LinHu06,LinHu07} and quantum entanglement  \cite{LCH08,LCH10,LinHu09,LinHu10,LCH15,LCH16,HHEnt}.}.   Therefore to see the dynamics with memory  
it is highly desirable to have  exact solutions (arbitrary coupling strength) to fully nonequilibrium (as opposed to linear response restricted by very weak coupling) dynamics. Since the harmonic oscillator detector linearly coupled to a quantum field is a Gaussian system we can in principle derive exact quantum stochastic equations for the  open (reduced) system under strong coupling strength, for low temperatures and find exact solutions to them \cite{Fleming}. 

\paragraph{Response function} 
Note that the detector response in our context is not the Wightman function of the field but what is measured in the detector, and more importantly, this is a dynamical response which includes the field's influence on the detector.  We do this with the influence functional \cite{if},  the closed-time-path   \cite{ctp} formalisms and  quantum Langevin equation \cite{QLE} techniques,  well adapted to the detector-field system.  We briefly summarize the key findings from these earlier work relevant to our present quests \cite{JH1,QRad,RHA,CPR} below.  

A nonMarkovian equation known as the Hu-Paz-Zhang (HPZ) master  equation for the  Brownian motion of a quantum harmonic oscillator   in a general environment has been derived in \cite{HPZ}. The HPZ master equation is an exact  nonMarkovian equation which preserves the positivity of the density operator and is valid for a) all temperatures, b) arbitrary spectral density of the bath, and c) arbitrary coupling strength between the system and the bath. Allowing for time-dependence in the natural frequency of the system  oscillator and in the $N$-oscillators making up its thermal bath, Hu and Matacz \cite{HM94} derived the master equations for the reduced density matrix of a parametric quantum oscillator system in a squeezed thermal bath and applied them to a range of problems: They showed that a detector sees \textit{thermal radiance} under uniform acceleration (Unruh), in a 2D black hole (Hartle-Hawking) and in a (static) de Sitter (Gibbons-Hawking) spacetime. Later, Raval, Koks, Hu and Matacz \cite{RHK97,KHMR}  showed that thermal radiation is emitted from a moving mirror and a collapsing shell, and near-thermal radiance is present in asymptotic conditions admitting exponential red-shifting.   

Here we continue to use the detector-field model to probe further into cosmological problems. 

\paragraph{Power spectrum} 
Besides the detector response a  commonly invoked quantity of physical interest is the two-point correlation function. The power spectrum is defined to be a suitable normalization factor times the spatial Fourier transform of the contracted correlation function at equal times. Dynamical responses as two-point functions are commonly discussed  in condensed matter physics \cite{Lovesey,Chaikin}. We focus only on cosmological problems here.  

In the two examples we mentioned earlier, the power spectrum of gravitational potential fluctuations is given by Eq. (28.67) and the correlation matrix by Eq. (28.69) of \cite{ThoBla}. The graviton two-point function is given by Eq. (4.34) of \cite{FRV12}, using the definition of the power spectrum defined in \cite{BFM92}. 
Those are, of course,  our witnesses $W$  
which co-existed and evolved with the quantum field throughout the entire cosmological history.  

\paragraph{What can a later time detector read about earlier history?}
In the second part of this paper we consider a detector introduced at the present time for cosmological observations, what we called the `Detective' $D$. We shall consider a harmonic oscillator of constant frequency, but the field modes have time-dependent frequencies for all times, getting excited by the expanding universe. Thus the squeeze parameter varies in time. To simplify the investigation, we study the frequency variation in a statically-bounded evolution with a valid `in-state, out-state'  description. At late times a stationarity  condition exists which makes the problem more tractable. The validity of a fluctuation-dissipation relation in the detector after relaxation is addressed in a companion paper \cite{FDRSq}. Under this condition we inquire to what extent the dynamical response of the detector introduced at a late time can read into the complex structure of the quantum field and reveal some earlier history of the universe.  

There are other important cosmological problems a detector-field system can help to investigate. Quantum decoherence via a single detector-quantum field model and quantum entanglement between two detectors in a common quantum field 
are two important topics which unfortunately we have no space to delve into in this paper. Suffice it to say that the conceptual framework of open quantum systems and nonequilibrium quantum field theory have finally entered the mainstream of current cosmological research. See, e.g., \cite{Fukuma,Boyan,Burgess,Nelson} and references therein.

This paper is organized as follows:  In Section~II, we briefly summarize field quantization in dynamical spacetimes and show that, under suitable transformation, the amplitude functions of the field modes behave like parametric oscillators. With this, we can use as our starting point a simple scalar field with time-dependent effective mass in flat space. In Sec.~III, we derive a  quantum equation of motion for the internal degree of freedom of an Unruh-DeWitt detector of fixed frequency coupled to a parametrically driven scalar field in Minkowski spacetime.  Solving this equation for a detector, assumed to have co-existed with the field in its evolution, will tell us the past history of the field squeezed by the universe's expansion. In Sec.~IV, we examine the features of this parametrically driven quantum field, and show how nonadiabaticity in the time-evolution of the field modes results in the squeezing and particle creation. These are the physical observables in the evolutionary history of the field  which we aim for a detector to pick up.  In. Sec.~V, we consider the case of a detector introduced at the present time  and investigate its internal dynamics to assess the potential of retrieving information about the past history of the field evolution. In Sec.~VI we summarize our findings for the two detector scenarios followed by some discussions.

\section{Quantum fields in dynamical spacetimes}
To make the presentation self-contained we give in this section a short summary of quantum field theory in curved space suitable for the considerations of cosmological processes mentioned earlier.  There are many reviews and monographs devoted to this subject, e.g., \cite{FullingBook,WaldBook,BirDav,ParToms,HuVer20}. Readers familiar with this subject can skip this section all together. We follow the treatment of \cite{CalHu08} here.  

\subsection{Free fields in flat space}
Some basic structure of quantum field theory (QFT) in flat space is needed for the construction of  QFT  in  curved spacetime.  
Consider a free massive $m$ scalar field  in Minkowski spacetime. The Heisenberg equation of motion for this field becomes the \textit{Klein-Gordon equation } ${\nabla }^{2}\hat{\Phi} (x)-m^{2}\hat{\Phi} (x)=0$. 
Assuming that the field lives in a finite large volume $V$ and expanding the scalar field operator in (spatial) Fourier modes, we have
\begin{equation}
	\hat{\Phi} (t,\bm {x})=\frac{1}{\sqrt{V}}\sum_{\bm {k}}\hat{\varphi} _{\bm {k}}(t){u}_{\bm {k}}(\bm {x})\,,  \label{modexp}
\end{equation}
where $\bm {k}=2\pi \bm {n}/L$, and $\bm {n}=(n_{1},n_{2},n_{3})$ in general consists of a triplet of integers. In Minkowski space the spatial mode functions are simply $u_{\bm {k}}=e^{i\bm {k}\cdot \bm {x}}$. In the infinite volume continuum limit this becomes
\begin{equation}
	\hat{\Phi} (\bm {x},t)=\int\!\frac{d^{3}\bm{k}}{( 2\pi)^{\frac{3}{2}}}\;e^{i\bm {k}\cdot\bm {x}}\hat{\varphi} _{\bm {k}}( t)  \label{d19a1}
\end{equation}
The operator-valued function $\hat{\varphi} _{\bm {k}}(t)$ for each mode $\bm {k}$ obeys a harmonic oscillator equation
\begin{equation}
	\frac{d^{2}\hat{\varphi} _{\bm {k}}}{dt^{2}}+\omega _{\bm {k}}^{2}\hat{\varphi} _{\bm {k}}=0,  \label{paramphi}
\end{equation}
where $\omega _{\bm {k}}^{2}= |\bm {k}|^{2}+ m^{2}$ in Minkowski space.

Given two complex independent solutions $f_{\bm {k}}^{\vphantom{*}}$, $f_{\bm {k}}^{*}$ of Eq.~\eqn{paramphi}, we may write
\begin{equation}
	\hat{\varphi} _{\bm {k}}( t) =\hat{a}_{+\bm {k}}^{\vphantom{\dagger}}f^{\vphantom{\dagger}}_{\bm {k}}( t) +\hat{a}_{-\bm {k}}^{\dagger }f_{\bm {k}}^{*\vphantom{\dagger}}( t) \,.  \label{d19c}
\end{equation}
Let us introduce the   the inner product  product $(f,g)=f\dot{g}-g\dot{f}$ of the two functions $f$ and $g$, which is conserved by Eq.~\eqn{paramphi}, and impose the normalization
\begin{equation}
	( f_{\bm {k}}^{\vphantom{*}},\,f_{\bm {k}}^{*}) =i\, .  \label{wron}
\end{equation}
The equal-time commutation relations are equivalent to
\begin{align}
	\bigl[ \hat{a}_{\bm {k}}^{\vphantom{\dagger}},\hat{a}_{\bm {k}'}^{\vphantom{\dagger}}\bigr] =\bigl[ \hat{a}_{\bm {k}}^{\dagger},\hat{a}_{\bm {k}'}^{\dagger }\bigr] &=0\,,	&\bigl[\hat{a}_{\bm {k}}^{\vphantom{\dagger}},\hat{a}_{\bm {k}'}^{\dagger }\bigr]& =\delta^{(3)} ( \bm {k}-\bm {k}')\,.
\label{d19e}
\end{align}
The operators $\hat{a}_{\bm {k}}^{\vphantom{\dagger}}$, $\hat{a}_{\bm {k}}^{\dagger }$ may be interpreted as particle annihilation and creation operators. We say that each choice of the basis functions $f_{\bm {k}}$ constitutes a \textit{particle model}\index{particle model}, where $f_{\bm {k}}$ is the \textit{positive frequency}\index{frequency!positive} component and $f_{\bm {k}}^{*}$ is the \textit{negative frequency}\index{frequency!negative} component of the $\bm {k}$-th mode; the state which is annihilated by all $a_{\bm {k}}$ is the vacuum of the particle model. The vacua of different particle models are in general inequivalent. This situation becomes more challenging for quantum fields in a dynamical background field or spacetime.



\subsection{QFT in curved spacetime}
Consider a massive scalar field $\Phi(x)$ of mass $m$ coupled arbitrarily $\xi$ to a background spacetime with metric $g_{\mu \nu}$ and scalar curvature $R$. Its dynamics is described by the action 
\begin{equation}
	S_{\Phi}=-\frac{1}{2}\int\!d^{4}x\sqrt{-g}\;\biggl\{g^{\mu \nu }(x)\nabla _{\mu }\Phi(x) \nabla _{\nu }\Phi(x) +\bigl[m^{2}+\xi R(x)\bigr]\Phi^{2}(x)\biggr\}\,,
\end{equation}
where the 0th  component of  $x^\mu$  denotes the time $\{t\}$ and the 1, 2, 3  the spatial $\{\bm{x}\}$ components,  with four-vector indices $\mu$, $\nu$ running from 0 to 3. (We may drop the $\mu$ index on $x^\mu$ for brevity).  The field-spacetime coupling constant $\xi=0$  corresponds    to minimal coupling and $\xi=1/6$ to conformal  coupling. Here, $\nabla_{\mu}$ denotes the covariant derivative with respect to the background spacetime with metric tensor $g_{\mu \nu }$ and $g=\det g_{\mu \nu }$ is its determinant.   The scalar field satisfies the wave equation
\begin{equation}
	\Bigl[\square-m^{2}-\xi R(x)\Bigr]\Phi (x)=0\,,  \label{E2}
\end{equation}
where the Laplace-Beltrami operator $\square$   defined on the background spacetime is given by,
\begin{equation}
	\square= g^{\mu \nu }\nabla _{\mu }\nabla _{\nu }=\frac{1}{\sqrt{-g(x)}}\frac{\partial }{\partial x^{\mu }}\left[ g^{\mu \nu }(x)\sqrt{-g(x)}\,\frac{\partial }{\partial x^{\nu }}\right]\,.
\end{equation}
In flat space, Poincar\'{e} invariance ensures the existence of a unique global Killing vector $\partial _{t}$ orthogonal to all constant-time spacelike hypersurfaces, an unambiguous separation of the positive-and negative-frequency modes, and a unique and well-defined vacuum. In curved spacetime, general covariance precludes any such privileged choice of time and slicing. There is no natural mode decomposition and no unique vacuum \cite{Fulling73,FullingBook}. We assume the background spacetime under consideration has at least enough symmetry to allow for a normal mode decomposition of the invariant operator at any constant-time slice.

In the canonical quantization approach, the quantum field $\Phi$ and its conjugate momentum $\Pi$ obey the equal time
commutation relation 
\begin{equation}
	\bigl[ \Phi (\bm {x},t) ,\Pi(\bm {x}',t)\bigr]=i\, \delta^{(3)}(\bm {x},\bm {x}')
\end{equation}
Here the scalar delta function $\delta^{(3)}(\bm {x},\bm {x}')$ in curved spacetime satisfies 
\begin{equation}
	\int\!d^{3}\bm{x}'\sqrt{-h(\bm{x}')}\;\delta^{(3)} (\bm {x},\bm {x}')\,F(\bm {x}')=F(\bm {x})\,, 
\end{equation}	
where $h$ is the determinant of the 3-metric on the spacelike surface and $F$ is any well behaved test function.

In the spatially flat Friedmann-Lemaitre-Robertson-Walker (FLRW) spacetime with line element  
\begin{equation}
	ds^{2}=-dt^{2}+a^{2}(t)\,d\bm{x}^{2}\,,
\end{equation}
the spatial mode functions are $u_{\bm {k}}=e^{i\bm {k}\cdot\bm {x}}$ and the wave equation for the amplitude function $f_{\bm {k}}(t)$ of mode $\bm {k}$ in cosmic time $t$ becomes 
\begin{equation}
	\ddot{f}_{k}(t)+3H\;\dot{f}_{k}(t)+\bigl[\omega^{2}(t)+\xi R(t)\bigr]f_{k}(t)=0\,,\label{costweq}
\end{equation}
where an overdot denotes the derivative with respect to cosmic time, and because of spatial isotropy, $f_{\bm {k}}$ in fact is a function of $k=\lvert\bm{k}\rvert$. In addition we introduce
\begin{align}
	\omega^{2}(t)&=\frac{k^{2}}{a^{2}(t)}+m^{2}\,,&R&=6\bigl[ \dot{H}(t)+{2}H^{2}(t)\bigr]\, ,  \label{qk}
\end{align}
with $H(t)=\dot{a}/a$ being the Hubble expansion rate of the background space.

The inequivalence of Fock representation in curved space, originating from the absence of a global time-like Killing vector, leads to an unavoidable mixing of positive- and negative-frequency components of the field, and then particle creation. In cosmological spacetimes the vacua defined at different times of evolution are not equivalent, so cosmological particle creation is by nature a dynamically induced effect. We stress that the particles in this context are not produced by interactions; they are excitations of the free-field by the changing background spacetimes. The physical mechanism is also different from thermal particle creation in black holes \cite{Haw74}, accelerated detectors \cite{Unr76} or moving mirrors \cite{FulDav76,DavFul77}, which involves the presence of a horizon, but underlies  dynamical Casimir effect \cite{DCE}. 

Unlike for QFT in constant background field or static spacetimes, a consistent separation into positive and negative energy solutions of the wave equation is not always possible. The definition of a vacuum\index{vacuum} state becomes a fundamental challenge in the construction of QFT in time-dependent backgrounds \cite{Fulling73}.
There are a few situations where vacuum states in QFT in dynamical spacetimes are well-defined: 1) the so-called statically bounded or asymptotically stationary spacetimes, where it is assumed that at $t=\pm \infty $ the background spacetime becomes stationary and the background fields become constant; 2) conformally-invariant fields in conformally static spacetimes. In both cases the Fock spaces are well defined and one can calculate the amplitude for particle creation  in an $S$-matrix sense. 3) If the background spacetime does not change too rapidly (quantified by a nonadiabaticity parameter described below) there is a conceptually clear and technically simple method in defining the so-called ($n$th order) \textit{adiabatic vacuum or number state}.

We shall consider cases 1) and 2) here for simplicity and bypass Case 3) where their treatments can be found in papers since the 70s.   We shall leave the topic of Bogoliubov transformation and particle creation, adiabatic number states and adiabatic regularization of the stress tensor  for the reader to consult the original papers.  

\subsection{Conformally-related spacetimes}
The spatially-flat FLRW spacetime $g_{\mu\nu}$ can be conformally related to the Minkowski spacetimes $\eta _{\mu \nu }$ by
\begin{align}
	g_{\mu \nu }(x)&=a^{2}(\eta )\eta _{\mu \nu }\,,&ds^{2}&=a^{2}(\eta )(-d\eta ^{2}+d\bm {x}^{2})\,,
\end{align}
where the conformal time $\eta$ is defined by
\begin{equation}	
	\eta=\int^{t}\!\frac{ds}{a(s)}\,,
\end{equation}
Since the conformally-related spacetime is static,   a global  Killing vector $\partial _{\eta }$ exists which enables us to define a globally well-defined vacuum state. The vacuum defined by the mode decomposition with respect to $\partial _{\eta }$ is known as the conformal vacuum.  For conformally-invariant fields, that is, a massless scalar field with $\xi =1/6$ in Eq.~\eqref{E2} in conformally-static spacetimes, there is no particle creation~\cite{Par72}. Any deviation from these conditions may result in particle production.

Consider a real massive scalar field coupled to a spatially-flat FLRW metric with constant coupling $\xi $. It is convenient to introduce a conformal amplitude function $\chi _{\bf k}(\eta )\equiv a(\eta)f_{\bf  k}(\eta )$ by rescaling the amplitude function $f_{\bf k}(t)$ in \eqref{costweq} for the normal mode ${\bf k}$. It then satisfies
the equation of motion of the parametric oscillator,
\begin{equation}
	\chi''_{k}(\eta)+\omega^{2}_c(\eta)\,\chi_{k}(\eta)=0\,,  \label{contweq}
\end{equation}
where a prime denotes $d/d\eta $, $k \equiv | {\bf k}|$ and
\begin{equation}
	\omega^{2}_c(\eta )=\omega^{2}(t)a^{2}(\eta)=k^{2}+a^{2}(\eta)\Bigl[m^{2}+(\xi -\frac{1}{6})\,R(\eta)\Bigr]\,,\label{Omega}
\end{equation}
is the conformal time-dependent frequency, with a subscript $c$ indicating that it is a conformally-related quantity.

One sees that, for massless ($m=0$) conformally coupled ($\xi=\frac{1}{6}$) fields in a spatially flat FLRW universe, the conformal wave equation admits solutions
\begin{equation}
	\chi_{k}(\eta )=A\,e^{+i\omega_c\eta }+B\,e^{-i\omega_c\eta}\,,
\end{equation}
which are of the same form as traveling waves in flat space. Since for $m=0$, $\xi=1/6$, we have a constant frequency $\omega=k$, the positive- and negative-frequency components remain separated and there is no particle production.

{Take as another example, the minimally coupled ($\xi=0$), massless scalar field in the spatially-flat de Sitter universe (the Poincar\'e patch) with natural frequency
\begin{equation}
	\omega_c^{2}(\eta)=k^{2}-\frac{2}{\eta^{2}}\,.
\end{equation}
The solutions to the wave equations for the amplitudes of the normal modes are given by
\begin{equation}
	\chi_{k}(\eta )=A\,e^{+ik\eta}\Bigl(1+\frac{i}{k\eta}\Bigr)+B\,e^{-ik\eta}\Bigl(1-\frac{i}{k\eta}\Bigr)\,.
\end{equation}
We see  a frozen power spectrum of the field on the superhorizon scale $k\eta\ll1$.   This example illustrates an important fact that the quantum field can have distinct behavior over various scales in the frequency modulation,   exhibited through different stages of  the field's evolution. In this work we look into the possibility of identifying these features imprinted in the nonMarkovian dynamics of one or more detectors that measure/interact with the quantum field.}

{The frequency modulation \eqref{Omega} captures the essence of the quantum fields in many interesting cosmological spacetimes which can be conformally-related  to that  in flat space.  We shall henceforth focus on parametric quantum fields in flat space, establishing a generic model for the investigation of nonMarkovianity in the dynamics of the detector and the changing field, then explore features that may allow us to fathom the past history of the Universe.}

\section{Quantum oscillator in a parametric quantum field}

Using the model of~\cite{HM94} we study the nonequilibrium dynamics of a Unruh-DeWitt detector coupled to a quantum field $\Phi(\bm{x}, t)$ in a cosmological spacetime. {The expansion of the universe acts like a drive which changes the natural frequencies of the normal modes  of the quantum field like in a parametric oscillator. Similarly for the quantum field driven by a moving conducting plate, as in the dynamical Casimir effect \cite{DCE}.  We shall call a driven quantum field with time-varying normal-mode natural frequency  a `parametric field'  for short. (Note that an undriven  quantum field has {a trivial} time dependence  $e^{\pm i \omega t}$.)     After the conformal transformation outlined in the previous section, its amplitude has a generic form like what  the amplitudes of a parametric field obey in flat space. Hereafter,   we shall treat a  real quantum scalar field $\Phi(\bm{x}, t)$ parametrically driven and evolving  in Minkowski space.

The action for the dynamics of the internal degrees of freedom $Q_{n}$ of the $n^{\text{th}}$ quantum parametric oscillator is given by
\begin{equation}
	S_{\textsc{udw}}=\int\!d^{3}\bm{x}\!\int\!dt\;\sum_{n}\frac{M}{2}\Bigl[\dot{Q}_{n}^{2}(t)-\Omega_{b}^{2}(t)\,Q_{n}^{2}(t)\Bigr]\,\delta^{(3)}(\bm{x}-\bm{z}_{n})\,,
\end{equation}
of mass $M$ and the bare oscillating frequency $\Omega_{b}$, and the detector's external degree of freedom follows a prescribed trajectory $\bm{z}_{n}(t)$. The quantum scalar field $\Phi(\bm{x},t)$ has a time-dependent mass $\mathfrak{m}(t)$, which may accommodate the effects from the external time-dependent sources, such as $a(t)$, that accounts for the parametric process of the field. Its action takes the form 
\begin{equation}\label{E:dgksjfbrt}
	S_{\textsc{b}}=\int\!d^{3}\bm{x}\!\int\!dt\;\frac{1}{2}\biggl\{\Bigl[\partial_{t}\Phi(\bm{x},t)\Bigr]^{2}-\Bigl[\nabla_{\bm{x}}\Phi(\bm{x},t)\Bigr]^{2}-\mathfrak{m}^{2}(t)\,\Phi^{2}(\bm{x},t)\biggr\}\,.
\end{equation}
The interaction between the detector and the field takes a simple bilinear form
\begin{equation}
	S_{\textsc{int}}=\int\!d^{3}\bm{x}\!\int\!dt\;\sum_{n}e(t)Q_{n}(t)\delta^{(3)}(\bm{x}-\bm{z}_{n})\,\Phi(\bm{x},t)\,,
\end{equation}
in which the detector-field coupling $e(t)$ is allowed to be time-dependent to include additional effects of the external driving sources.

These actions lead to the following coupled Heisenberg equations for the internal degrees $\hat{Q}_{n}(t)$ of freedom of the detectors and the field $\hat{\Phi}(\bm{x},t)$,
\begin{align}
	M\,\ddot{\hat{Q}}_{n}(t)+M\Omega_{b}^{2}(t)\,\hat{Q}_{n}(t)&=e(t)\,\hat{\Phi}(\bm{z}_{n},t)\,,\label{E:ihtdnkj1}\\
	\frac{\partial^{2}}{\partial t^{2}}\hat{\Phi}(\bm{x},t)-\nabla^{2}_{\bm{x}}\hat{\Phi}(\bm{x},t)+\mathfrak{m}^{2}(t)\,\hat{\Phi}(\bm{x},t)&=\sum_{n}e(t)Q_{n}(t)\delta^{(3)}(\bm{x}-\bm{z}_{n})\,.\label{E:ihtdnkj2}
\end{align}
Formally solving \eqref{E:ihtdnkj2} gives
\begin{align}
	\hat{\Phi}(\bm{x},t)&=\hat{\Phi}_{h}(\bm{x},t)+\int\!d^{3}\bm{x}'ds\;G_{R,0}^{(\Phi)}(\bm{x},t;\bm{x}',s)\sum_{n}e(s)Q_{n}(s)\delta^{(3)}(\bm{x}'-\bm{z}_{n})\notag\\
	&=\hat{\Phi}_{h}(\bm{x},t)+\int_{0}^{t}\!ds\;\sum_{n}e(s)\,G_{R,0}^{(\Phi)}(\bm{x},t;\bm{z}_{n},s)Q_{n}(s)\,,\label{E:irnddfgd}
\end{align}
where $\hat{\Phi}_{h}(\bm{x},t)$ is the homogeneous solution to the wave equation \eqref{E:ihtdnkj2} and describes the free field, while $\hat{\Phi}(\bm{x},t)$ on the lefthand side of \eqref{E:irnddfgd}, which we call the interacting field, comprises the back-action of the internal degree of freedom of the detector in the form of radiation field \cite{QRad} as can be seen from the second term on the righthand side. This difference is subtle but worth noticing. 

The two-point function $G_{R,0}^{(\Phi)}(\bm{x},t;\bm{x}',t')$ is the retarded Green's function of the field, satisfying the inhomogeneous wave equation
\begin{equation*}
	\frac{\partial^{2}}{\partial t^{2}}G_{R,0}^{(\Phi)}(\bm{x},t;\bm{x}',t')-\nabla^{2}_{\bm{x}}G_{R,0}^{(\Phi)}(\bm{x},t;\bm{x}',t')+\mathfrak{m}^{2}(t)\,G_{R,0}^{(\Phi)}(\bm{x},t;\bm{x}',t')=\delta^{3}(\bm{x}-\bm{x}')\delta(t-t')\,.
\end{equation*}
Plugging \eqref{E:irnddfgd} back to \eqref{E:ihtdnkj1} thus yields  
\begin{equation}\label{E:gbksddf}
	M\,\ddot{\hat{Q}}_{n}(t)+M\Omega_{b}^{2}(t)\,\hat{Q}_{n}(t)=e(t)\hat{\Phi}_{h}(\bm{z}_{n},t)+\sum_{j}\int_{0}^{t}\!ds\;e(t)G_{R,0}^{(\Phi)}(\bm{z}_{n},t;\bm{z}_{j},s)e(s)Q_{j}(s)\,.
\end{equation}
This governs the nonequilibrium dynamics of the internal degree of freedom of the detector, coupled to a parametric field. The expressions on the righthand side result from the detector-field interaction. The first term describes the fluctuating force due to the quantum fluctuations of the free field. The second term is a nonlocal expression, which contains 1) a local frictional force, 2) a {\it self-non-Markovian effect} of the detector due to the finite effective mass of the field, and 3) {\it mutual non-Markovian influences} between detectors, mediated by the field. We will dwell on these points  after we have derived the field dynamics.

\section{Dynamics of parametrically-driven quantum field}

Suppose the parametric process of the free classical field starts at $t=\mathfrak{t}_{i}>0$ and ends at $\mathfrak{t}_{f}$, during which the effective mass $\mathfrak{m}(t)$ increases monotonically and smoothly from one fixed value $\mathfrak{m}_{i}$ to another $\mathfrak{m}_{f}>\mathfrak{m}_{i}$. The duration of the parametric process is then $\mathfrak{t}=\mathfrak{t}_{f}-\mathfrak{t}_{i}$. Thus before $\mathfrak{t}_{i}$ and after $\mathfrak{t}_{f}$, the field $\Phi(\bm{x},t)$ behaves like a free, real massive  scalar field. Note that in this section we suppress the subscript $h$ for the free field. We will put the subscript back when needed.

Let us expand $\Phi(\bm{x},t)$ by
\begin{equation}
	\Phi(\bm{x},t)=\int\!\!\frac{d^{3}\bm{k}}{(2\pi)^{\frac{3}{2}}}\;e^{+i\bm{k}\cdot\bm{x}}\,\varphi_{\bm{k}}(t)\,,
\end{equation}
with the mode function $\varphi_{\bm{k}}^{*}(t)=\varphi_{-\bm{k}}^{\vphantom{*}}(t)$. The action of the free field \eqref{E:dgksjfbrt} then takes the form of the action for a collection of parametric oscillators
\begin{equation}
	S=\int\!dt\;\frac{1}{2}\int\!d^{3}\bm{k}\,\Bigl\{\dot{\varphi}_{\bm{k}}^{\vphantom{*}}(t)\dot{\varphi}_{\bm{k}}^{*}(t)-\omega^{2}(t)\,\varphi_{\bm{k}}^{\vphantom{*}}(t)\varphi_{\bm{k}}^{*}(t)\Bigr\}\,,
\end{equation}
with natural frequency $\omega(t)$ 
\begin{equation}\label{E:kgfbdf}
	\omega^{2}(t)=\bm{k}^{2}+\mathfrak{m}^{2}(t)\,,
\end{equation}
such that the frequency monotonically and smoothly changes from $\omega_{i}$ at the beginning of the parametric process to $\omega_{f}$ at the end
\begin{align}
	\omega_{i}^{2}&=\bm{k}^{2}+\mathfrak{m}_{i}^{2}\,,&\omega_{f}^{2}&=\bm{k}^{2}+\mathfrak{m}_{f}^{2}\,.
\end{align}
The mode function $\varphi_{\bm{k}}(t)$ satisfies the equation of motion of a classical parametric oscillator
\begin{equation}\label{E:rirtyrd}
	\ddot{\varphi}_{\bm{k}}(t)+\omega^{2}(t)\,\varphi_{\bm{k}}(t)=0\,.
\end{equation}
Its solutions of the corresponding Heisenberg equation can be formally given in terms of the initial conditions at $t=0<\mathfrak{t}_{i}$
\begin{align}
	\hat{\varphi}_{\bm{k}}(t)&=d^{(1)}_{\bm{k}}(t)\,\hat{\varphi}_{\bm{k}}(0)+d^{(2)}_{\bm{k}}(t)\,\hat{\pi}_{-\bm{k}}(0)\,,\label{E:dgbksbgsd1}\\
	\hat{\pi}_{\bm{k}}(t)&=\dot{d}^{(1)}_{-\bm{k}}(t)\,\hat{\varphi}_{-\bm{k}}(0)+\dot{d}^{(2)}_{-\bm{k}}(t)\,\hat{\pi}_{\bm{k}}(0)\,,
\end{align}
where $\hat{\pi}_{\bm{k}}=\dot{\varphi}_{-\bm{k}}(t)$ is the mode function of the canonical momentum $\hat{\Pi}$ conjugated to the field $\hat{\Phi}$. We introduce a special set of homogeneous solutions $d^{(1)}_{\bm{k}}(t)$, $d^{(2)}_{\bm{k}}(t)$ to \eqref{E:rirtyrd}, satisfying
\begin{align}
	d^{(1)}_{\bm{k}}(0)&=1\,,&\dot{d}^{(1)}_{\bm{k}}(0)&=0\,,&d^{(2)}_{\bm{k}}(0)&=0\,,&\dot{d}^{(2)}_{\bm{k}}(0)&=1\,,
\end{align}
for each mode $\bm{k}$. They are particularly convenient in the context of nonequilibrium dynamics. The canonical commutation relation $[\hat{\varphi}_{\bm{k}}(t),\hat{\pi}_{\bm{k}'}(t)]=i\,\delta_{\bm{k}\bm{k}'}$ gives
\begin{equation}\label{E:fgksbrt}
	d^{(1)}_{\bm{k}}(t)\dot{d}^{(2)}_{\bm{k}}(t)-d^{(2)}_{\bm{k}}(t)\dot{d}^{(1)}_{\bm{k}}(t)=1\,,
\end{equation}
which is nothing but the Wronskian corresponding to \eqref{E:rirtyrd}. We also note that $d^{(i)}_{\bm{k}}(t)$ only depends on the magnitude of $\bm{k}$ due to \eqref{E:kgfbdf}.

Now suppose at $t=0$, that is, $f_{\bm {k}}=1$ in \eqref{d19c}, we rewrite the field-mode operator and its momentum in terms of the creation and annihilation operators $\hat{a}^{\vphantom{\dagger}}_{\bm{k}}$, $\hat{a}^{\dagger}_{\bm{k}}$,
\begin{align}
	\hat{\varphi}_{\bm{k}}&=\frac{1}{\sqrt{2\omega_{i}}}\,\bigl(\hat{a}^{\dagger}_{\bm{k}}+\hat{a}^{\vphantom{\dagger}}_{\bm{k}}\bigr)\,,&\hat{\pi}_{\bm{k}}=i\sqrt{\frac{\omega_{i}}{2}}\,\bigl(\hat{a}^{\dagger}_{-\bm{k}}-\hat{a}^{\vphantom{\dagger}}_{-\bm{k}}\bigr)\,,
\end{align}
such that by \eqref{E:dgbksbgsd1}, we can express $\hat{\varphi}_{\bm{k}}(t)$ in terms of $\hat{a}^{\vphantom{\dagger}}_{\bm{k}}(0)$ and $\hat{a}^{\dagger}_{\bm{k}}(0)$ 
\begin{align}\label{E:rnbryjy}
	\hat{\varphi}_{\bm{k}}(t)=\frac{1}{\sqrt{2\omega_{i}}}\Bigl\{\Bigl[d^{(1)}_{\bm{k}}(t)-i\,\omega_{i}\,d^{(2)}_{\bm{k}}(t)\Bigr]\,\hat{a}^{\vphantom{\dagger}}_{\bm{k}}(0)+\Bigl[d^{(1)}_{\bm{k}}(t)+i\,\omega_{i}\,d^{(2)}_{\bm{k}}(t)\Bigr]\,\hat{a}^{\dagger}_{\bm{k}}(0)\Bigr\}\,,
\end{align}
with the shorthand notation $\omega(t)=\omega_{i}$ when $t\leq\mathfrak{t}_{i}$. Eq.~\eqref{E:rnbryjy} gives the formal, exact dynamics at any time $t>0$ of the free parametric field according to the action \eqref{E:dgksjfbrt} with given initial conditions at $t=0$. Then, the field operator $\hat{\Phi}(\bm{x},t)$ has a plane-wave expansion of the form
\begin{align}
	\hat{\Phi}(\bm{x},t)&=\int\!\!\frac{d^{3}\bm{k}}{(2\pi)^{\frac{3}{2}}}\frac{1}{\sqrt{2\omega_{i}}}\Bigl\{\Bigl[d^{(1)}_{\bm{k}}(t)-i\,\omega_{i}\,d^{(2)}_{\bm{k}}(t)\Bigr]\,\hat{a}^{\vphantom{\dagger}}_{\bm{k}}(0)\,e^{+i\bm{k}\cdot\bm{x}}\Bigr.\notag\\
	&\qquad\qquad\qquad\qquad\qquad+\Bigl.\Bigl[d^{(1)}_{\bm{k}}(t)+i\,\omega_{i}\,d^{(2)}_{\bm{k}}(t)\Bigr]\,\hat{a}^{\dagger}_{\bm{k}}(0)\,e^{-i\bm{k}\cdot\bm{x}}\Bigr\}\,,\label{E:ngkfjg1}
\end{align}
and the corresponding momentum operator is
\begin{align}
	\hat{\Pi}(\bm{x},t)&=\int\!\!\frac{d^{3}\bm{k}}{(2\pi)^{\frac{3}{2}}}\frac{1}{\sqrt{2\omega_{i}}}\Bigl\{\Bigl[\dot{d}^{(1)}_{\bm{k}}(t)-i\,\omega_{i}\,\dot{d}^{(2)}_{\bm{k}}(t)\Bigr]\,\hat{a}^{\vphantom{\dagger}}_{\bm{k}}(0)\,e^{+i\bm{k}\cdot\bm{x}}\Bigr.\notag\\
	&\qquad\qquad\qquad\qquad\qquad+\Bigl.\Bigl[\dot{d}^{(1)}_{\bm{k}}(t)+i\,\omega_{i}\,\dot{d}^{(2)}_{\bm{k}}(t)\Bigr]\,\hat{a}^{\dagger}_{\bm{k}}(0)\,e^{-i\bm{k}\cdot\bm{x}}\Bigr\}\,.\label{E:ngkfjg2}
\end{align}
It can be easily seen that for the standard Klein-Gordon field in the unbounded Minkowski space, we have $\omega(t)=\omega_{i}$, and
\begin{align}
	\dot{d}^{(1)}_{\bm{k}}(t)&=\cos\omega_{i} t\,,&\dot{d}^{(2)}_{\bm{k}}(t)&=\frac{1}{\omega_{i}}\,\sin\omega_{i} t\,,
\end{align}
and the expansions \eqref{E:ngkfjg1} and \eqref{E:ngkfjg2} revert to the conventional plane-wave forms we are familiar with.

With the field expansion \eqref{E:ngkfjg1}, we are ready to construct its Green's functions. The retarded Green's function $G_{R,0}^{(\Phi)}(\bm{x},t;\bm{x}',t')$ of the free field $\hat{\Phi}(\bm{x},t)$ is given by
\begin{align}\label{E:bfkfkrte}
	G_{R,0}^{(\Phi)}(\bm{x},t;\bm{x}',t')&=i\,\theta(t-t')\,\bigl[\hat{\Phi}(\bm{x},t),\hat{\Phi}(\bm{x}',t')\bigr]\notag\\
	&=-\theta(t-t')\int\!\!\frac{d^{3}\bm{k}}{(2\pi)^{3}}\;e^{+i\bm{k}\cdot(\bm{x}-\bm{x}')}\Bigl\{d^{(1)}_{\bm{k}}(t)d^{(2)}_{\bm{k}}(t')-d^{(2)}_{\bm{k}}(t)d^{(1)}_{\bm{k}}(t')\Bigr\}\,,
\end{align}
where $\theta(t)$ is the unit-step function. This is state-independent in the current setting, but in contrast to the cases~\cite{CPR,PRE18,HHEnt,LinHu07,LinHu06} we have treated, it is in general nonstationary due to the parametric process involved. {The terms in the curly brackets will not reduce to $d^{(2)}_{\bm{k}}(t-t')$.} However, for the scenario we are considering, it behaves like the standard retarded Green's function of the massive field, but with different masses $\mathfrak{m}$ in the regimes $t$, $t'\leq\mathfrak{t}_{i}$ and $t$, $t'\geq\mathfrak{t}_{f}$. In these two regimes, this retarded Green's function become stationary in time translation.

\begin{figure}
\centering
   \scalebox{0.4}{\includegraphics{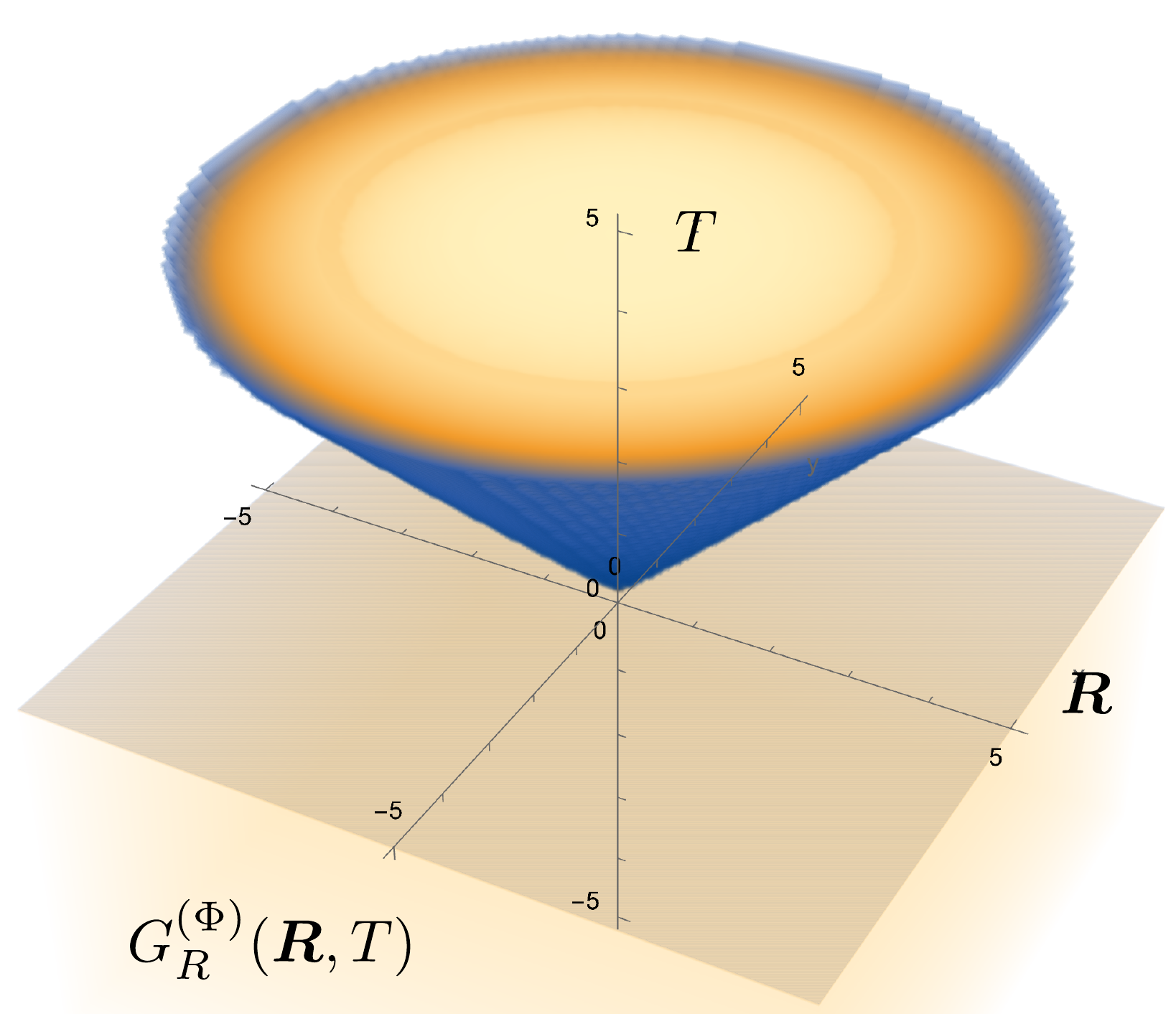}}
    \caption{The retarded Green's function of a massive Klein-Gordon field. Note that it has a non-zero contribution inside the forward lightcone.}\label{Fi:GrFnmm}
\end{figure}

The retarded Green's function and the Hadamard function of the free massive field of constant mass $m$ are given by~\cite{bjorken&drell}
\begin{align}
	G_{R,0}(x,x')&=\frac{\theta(T)}{2\pi}\,\Bigl[\delta(\sigma^{2})-\theta(\sigma^{2})\,\frac{m}{2\sqrt{\sigma^{2}}}\,J_{1}(m\sqrt{\sigma^{2}})\Bigr]\,,\label{E:kdfgbksfb1}\\
	G_{H,0}(x,x')&=\frac{1}{4\pi}\Bigl[\theta(+\sigma^{2})\,\frac{m}{2\sqrt{+\sigma^{2}}}\,Y_{1}(m\sqrt{+\sigma^{2}})+\theta(-\sigma^{2})\,\frac{m}{\pi\sqrt{-\sigma^{2}}}\,K_{1}(m\sqrt{-\sigma^{2}})\Bigr]\,,\notag
\end{align}
where $T=t-t'$ and the spacetime interval $\sigma^{2}=T^{2}-\bm{R}^{2}$ with $\bm{R}=\bm{x}-\bm{x}'$. In the case of a single detector at a fixed spatial location, we have $\bm{R}=0$, and hence $\sigma^{2}>0$ always. We see from \eqref{E:kdfgbksfb1} that other than the lightcone structure enforced by the first term on the righthand side, it has an additional term that does not vanish for $\sigma^{2}>0$, that is, in a timelike interval. It characterizes a duration of the memory of the order $m^{-1}$. This implies that the detector can affect its own subsequent motion by the radiation field it emits at the present time. In principle, the field has a longer memory for a lighter mass, but the small mass also reduces the ``strength'' of this {\it self-nonMarkovian} effect. Hence there is a compromise.

Next we turn to the Hadamard function $G_{H,0}^{(\Phi)}(\bm{x},t;\bm{x}',t')=\dfrac{1}{2}\,\langle\bigl\{\hat{\Phi}(\bm{x},t),\hat{\Phi}(\bm{x}',t')\bigr\}\rangle$, that is, the noise kernel of the field. It reflects the correlation of the field and provides information about the magnitude of the noise force on the detector. From the mode expansion \eqref{E:ngkfjg1}, we have
\begin{align}
	&\quad G_{H,0}^{(\Phi)}(\bm{x},t;\bm{x}',t')\notag\\
	&=\int\!\!\frac{d^{3}\bm{k}}{(2\pi)^{3}}\frac{1}{2\omega_{i}}\biggl\{\,e^{+i\bm{k}(\bm{x}-\bm{x}')}\Bigl(\langle\hat{a}_{\bm{k}}^{\dagger}(0)\hat{a}_{\bm{k}}^{\vphantom{\dagger}}(0)\rangle+\frac{1}{2}\Bigr)\Bigl[2d^{(1)}_{\bm{k}}(t)d^{(1)}_{\bm{k}}(t')+2\omega_{i}^{2}d^{(2)}_{\bm{k}}(t)d^{(2)}_{\bm{k}}(t')\Bigr]\biggr.\notag\\
	&\qquad\qquad\qquad\;+e^{+i\bm{k}(\bm{x}+\bm{x}')}\,\langle\hat{a}_{\bm{k}}^{\hphantom{\dagger}2}(0)\rangle\Bigl[d^{(1)}_{\bm{k}}(t)d^{(1)}_{\bm{k}}(t')-\omega_{i}^{2}\,d^{(2)}_{\bm{k}}(t)d^{(2)}_{\bm{k}}(t')\Bigr.\notag\\
	&\qquad\qquad\qquad\qquad\qquad\qquad\qquad\qquad\qquad-\Bigl.i\,\omega_{i}\,d^{(1)}_{\bm{k}}(t)d^{(2)}_{\bm{k}}(t')-i\,\omega_{i}\,d^{(2)}_{\bm{k}}(t)d^{(1)}_{\bm{k}}(t')\Bigr]\notag\\
	&\qquad\qquad\qquad\;+\biggl.e^{-i\bm{k}(\bm{x}+\bm{x}')}\,\langle\hat{a}_{\bm{k}}^{\dagger2}(0)\rangle\Bigl[d^{(1)}_{\bm{k}}(t)d^{(1)}_{\bm{k}}(t')-\omega_{i}^{2}\,d^{(2)}_{\bm{k}}(t)d^{(2)}_{\bm{k}}(t')\Bigr.\label{E:brtyufb}\\
	&\qquad\qquad\qquad\qquad\qquad\qquad\qquad\qquad\qquad+\Bigl.i\,\omega_{i}\,d^{(1)}_{\bm{k}}(t)d^{(2)}_{\bm{k}}(t')+i\,\omega_{i}\,d^{(2)}_{\bm{k}}(t)d^{(1)}_{\bm{k}}(t')\Bigr]\biggr\}\,.\notag
\end{align}
Here $\langle\cdots\rangle$ is the expectation value taken with respect to the initial state of the field at $t=0$. Thus if the initial state is a stationary state, then the last two lines vanish. Note that the operators inside the expectation values are evaluated at the initial time.

Since the Hadamard function is state-dependent, we need to specify the configuration of interest in more details before proceeding. Suppose we are interested in the  {regime} $t_{i}>\mathfrak{t}_{f}$, after the parametric process of the field ceases. It is known  {and shown below} that the state of the oscillator before the parametric process will be squeezed and rotated at the end of the process. This implies that, for the field which is a collection of parametric oscillators, the squeezing and rotation will be mode-dependent, so at the end of the parametric process, the state of the field is squeezed by a two-mode squeeze operator $\hat{S}_{2}(\zeta_{\bm{k}})$, {where} the squeeze parameter $\zeta_{\bm{k}}$ takes a polar decomposition, $\zeta_{\bm{k}}=\eta_{\bm{k}}\,e^{+i\theta_{\bm{k}}}$ with $\eta_{\bm{k}}\geq0$ and $0\leq\theta_{\bm{k}}<2\pi$. For the moment let's neglect the effect of rotation, because it can be absorbed into the squeeze angle $\theta_{\bm{k}}$ if necessary. We assume that at the initial time $t=0$ before the parametric process, the field is in a thermal state $\hat{\rho}_{\beta}^{(\Phi)}$. Then after the process the field will be in a squeezed thermal state
\begin{equation}
	\hat{\rho}_{\textsc{st}}^{(\Phi)}=\prod_{\bm{k}}\hat{S}_{2}^{\vphantom{\dagger}}(\zeta_{\bm{k}})\,\hat{\rho}_{\beta}^{(\Phi)}(\omega_{i})\,\hat{S}^{\dagger}_{2}(\zeta_{\bm{k}})\,,\label{E:ngritris}
\end{equation}
where the two-mode squeeze operator $\hat{S}_{2}(\zeta_{\bm{k}})$ is given by
\begin{equation}
	\hat{S}_{2}(\zeta_{\bm{k}})=\exp\Bigl[\zeta_{\bm{k}}^{*\vphantom{\dagger}}\hat{a}_{\bm{k}}^{\vphantom{\dagger}}(0)\hat{a}_{\shortminus\bm{k}}^{\vphantom{\dagger}}(0)-\zeta_{\bm{k}}^{\vphantom{\dagger}}\hat{a}_{\bm{k}}^{\dagger}(0)\hat{a}_{\shortminus\bm{k}}^{\dagger}(0)\Bigr]\,.
\end{equation}
The product in \eqref{E:ngritris} is understood as a shorthand notation that the squeeze  operators $\hat{S}_{2}(\zeta_{\bm{k}})$ are applied onto $\hat{\rho}_{\beta}^{(\Phi)}$ mode-by-mode wise. The squeeze operators of different $\bm{k}$ are orthogonal to one another. The thermal state $\hat{\rho}_{\beta}^{(\Phi)}(\omega_{i})$ is of the form
\begin{align}\label{E:fgkjdsfg}
	\hat{\rho}_{\beta}^{(\Phi)}(\omega_{i})&=\frac{1}{Z_{\beta}}\,\prod_{\bm{k}}\exp\Bigl[-\beta\bigl(n_{\bm{k}}+\frac{1}{2}\bigr)\,\omega_{i}\Bigr]\,\lvert n_{\bm{k}}\rangle\langle n_{\bm{k}}\rvert\,,&\omega_{i}&=\sqrt{\smash[b]{\lvert\bm{k}\rvert_{\vphantom{i}}^{2}+\mathfrak{m}_{i}^{2}}}\,,
\end{align}
where $Z_{\beta}$ is the corresponding partition function, $Z_{\beta}=\operatorname{Tr}_{\Phi}\hat{\rho}_{\beta}^{(\Phi)}(\omega_{i})$.

We can also place the effects of squeezing onto the field operator, and treat the field \eqref{E:brtyufb} as the consequences of the squeeze operator acting  on the \textit{in}-modes, that is, $\hat{S}_{2}^{\dagger}(\zeta)\,\hat{\Phi}_{\textsc{in}}(\bm{x},t)\,\hat{S}_{2}(\zeta)$ with
\begin{equation}\label{E:gkrbfkdf}
	\hat{\Phi}_{\textsc{in}}(\bm{x},t)=\int\!\frac{d^{3}\bm{x}}{(2\pi)^{\frac{3}{2}}}\;\frac{1}{\sqrt{2\omega_{i}}}\Bigl[\hat{a}_{\bm{k}}^{\vphantom{\dagger}}(0)\,e^{+i\bm{k}\cdot\bm{x}-i\omega_{i}t}+\hat{a}_{\bm{k}}^{\dagger}(0)\,e^{-i\bm{k}\cdot\bm{x}+i\omega_{i}t}\Bigr]\,.
\end{equation}
so that at any intermediate times, we have
\begin{align*}
	\hat{\Phi}(\bm{x},t)&=\int\!\!\frac{d^{3}\bm{k}}{(2\pi)^{\frac{3}{2}}}\;\frac{1}{\sqrt{2\omega_{i}}}\Bigl\{\Bigl[d^{(1)}_{\bm{k}}(t)-i\,\omega_{i}\,d^{(2)}_{\bm{k}}(t)\Bigr]\,\hat{a}^{\vphantom{\dagger}}_{\bm{k}}(0)\,e^{+i\bm{k}\cdot\bm{x}}\Bigr.\notag\\
	&\qquad\qquad\qquad\qquad\qquad\qquad\qquad\qquad+\Bigl.\Bigl[d^{(1)}_{\bm{k}}(t)+i\,\omega_{i}\,d^{(2)}_{\bm{k}}(t)\Bigr]\,\hat{a}^{\dagger}_{\bm{k}}(0)\,e^{-i\bm{k}\cdot\bm{x}}\Bigr\}\notag\\
	&=\int\!\!\frac{d^{3}\bm{k}}{(2\pi)^{\frac{3}{2}}}\;\frac{1}{\sqrt{2\omega_{i}}}\Bigl\{e^{-i\omega_{i}t}\,\hat{S}^{\dagger}(\zeta_{\bm{k}})\,\hat{a}^{\vphantom{\dagger}}_{\bm{k}}(0)\,\hat{S}_{2}(\zeta_{\bm{k}})\,e^{+i\bm{k}\cdot\bm{x}}\Bigr.\notag\\
	&\qquad\qquad\qquad\qquad\qquad\qquad\qquad\qquad+\Bigl.e^{+i\omega_{i}t}\,\hat{S}_{2}^{\dagger}(\zeta_{\bm{k}})\,\hat{a}^{\dagger}_{\bm{k}}(0)\,\hat{S}(\zeta_{\bm{k}})\,e^{-i\bm{k}\cdot\bm{x}}\Bigr\}\,.
\end{align*}
As such, we have
\begin{align*}
	G_{H,0}^{(\Phi)}(\bm{x},t;\bm{x}',t')=\frac{1}{2}\operatorname{Tr}_{\Phi}\biggl[\hat{\rho}^{(\Phi)}_{\beta}\bigl\{\hat{\Phi}(\bm{x},t),\hat{\Phi}(\bm{x}',t')\bigr\}\biggr]=\frac{1}{2}\operatorname{Tr}_{\Phi}\biggl[\hat{\rho}^{(\Phi)}_{\textsc{st}}\bigl\{\hat{\Phi}_{\textsc{in}}(\bm{x},t),\hat{\Phi}_{\textsc{in}}(\bm{x}',t')\bigr\}\biggr]\,.
\end{align*}
If we let
\begin{equation}\
	\hat{S}_{2}^{\dagger}(\zeta_{\bm{k}})\,\hat{a}_{\bm{k}}(0)\,\hat{S}_{2}^{\vphantom{\dagger}}(\zeta_{\bm{k}})=\alpha_{\bm{k}}^{\vphantom{*}}\,\hat{a}^{\vphantom{\dagger}}_{\bm{k}}(0)+\beta_{\shortminus\bm{k}}^{*}\,\hat{a}^{\dagger}_{\shortminus\bm{k}}(0)\,,\label{E:irgubdf}
\end{equation}
with $\alpha_{\bm{k}}^{\vphantom{*}}=\cosh\eta_{\bm{k}}^{\vphantom{*}}$, $\beta_{\bm{k}}^{\vphantom{*}}=-e^{-i\theta_{\bm{k}}^{\vphantom{*}}}\,\sinh\eta_{\bm{k}}^{\vphantom{*}}$, then we find
\begin{equation}\label{E:fnljsb1}
	d^{(1)}_{\bm{k}}(t)-i\,\omega_{i}\,d^{(2)}_{\bm{k}}(t)=e^{-i\omega_{i}t}\,\alpha_{\bm{k}}^{\vphantom{*}}(t)+e^{+i\omega_{i}t}\,\beta_{\bm{k}}^{\vphantom{*}}(t)\,.
\end{equation}
Similarly, for the conjugate momentum $\hat{\Pi}(\bm{x},t)$, we also find
\begin{equation}\label{E:fnljsb2}
	\dot{d}^{(1)}_{\bm{k}}(t)-i\,\omega_{i}\,\dot{d}^{(2)}_{\bm{k}}(t)=-i\,\omega_{i}\,e^{-i\omega_{i}t}\,\alpha_{\bm{k}}(t)+i\,\omega_{i}\,e^{+i\omega_{i}t}\,\beta_{\bm{k}}(t)\,.
\end{equation}
Eqs.~\eqref{E:fnljsb1} and \eqref{E:fnljsb2} lead to
\begin{align}
	\alpha_{\bm{k}}(t)&=\frac{1}{2\omega_{i}}\,e^{+i\omega_{i}t}\,\Bigl[\omega_{i}\,d^{(1)}_{\bm{k}}(t)+i\,\dot{d}^{(1)}_{\bm{k}}(t)-i\,\omega_{i}^{2}d^{(2)}_{\bm{k}}(t)+\omega_{i}\,\dot{d}^{(2)}_{\bm{k}}(t)\Bigr]\,,\label{E:gbkrjbkdg1}\\
	\beta_{\bm{k}}(t)&=\frac{1}{2\omega_{i}}\,e^{-i\omega_{i}t}\,\Bigl[\omega_{i}\,d^{(1)}_{\bm{k}}(t)-i\,\dot{d}^{(1)}_{\bm{k}}(t)-i\,\omega_{i}^{2}d^{(2)}_{\bm{k}}(t)-\omega_{i}\,\dot{d}^{(2)}_{\bm{k}}(t)\Bigr]\,.\label{E:gbkrjbkdg2}
\end{align}
We can verify that 
\begin{equation}
	\lvert\alpha_{\bm{k}}\rvert^{2}-\lvert\beta_{\bm{k}}\rvert^{2}=d^{(1)}_{\bm{k}}(t)\dot{d}^{(2)}_{\bm{k}}(t)-\dot{d}^{(1)}_{\bm{k}}(t)d^{(2)}_{\bm{k}}(t)=1\,,
\end{equation}
The righthand side of \eqref{E:irgubdf} gives the Bogoliubov transformation of the creation and annihilation operators at initial time, and the time-dependent Bogoliubov coefficients $\alpha_{\bm{k}}(t)$ and $\beta_{\bm{k}}(t)$ contain the information of the parametric processes of the field, which is encapsulated in the squeeze parameter $\zeta_{\bm{k}}(t)$. In particular, the factor $\lvert\beta_{\bm{k}}(t)\rvert^{2}$ gives the number density of particle production of the field quanta during the parametric amplification process. To see this better, let us define $\hat{b}_{\bm{k}}=\hat{S}_{2}^{\dagger}(\zeta_{\bm{k}})\,\hat{a}_{\bm{k}}(0)\,\hat{S}_{2}^{\vphantom{\dagger}}(\zeta_{\bm{k}}$. If the field is initially in the vacuum state $\lvert0_{a}\rangle$ defined by $\hat{a}_{\bm{k}}(0)$, namely, $\hat{a}_{\bm{k}}(0)\lvert0_{a}\rangle=0$, then we can readily show that 
\begin{equation}\label{E:nvkdjfsd}
	\sum_{\bm{k}}\langle0_{a}\vert\hat{b}_{\bm{k}}^{\dagger}\hat{b}_{\bm{k}}^{\vphantom{\dagger}}\vert0_{a}\rangle=\sum_{\bm{k}}\lvert\beta_{\bm{k}}\rvert^{2}\,.
\end{equation}
That is, the resulting two-mode squeezed vacuum has a nonvanishing number of particles, given by \eqref{E:nvkdjfsd}, compared to the initial vacuum. If the initial number state $\lvert n_{a}\rangle$ of the field is not a vacuum, and but has nonzero particle numbers $N_{\bm{k}}^{(a)\vphantom{\dagger}}=\langle n_{a}\vert\hat{a}_{\bm{k}}^{\dagger}(0)\hat{a}_{\bm{k}}^{\vphantom{\dagger}}(0)\vert n_{a}\rangle\neq0$, then we find
\begin{equation}
	\sum_{\bm{k}}\langle n_{a}\vert\hat{b}_{\bm{k}}^{\dagger}\hat{b}_{\bm{k}}^{\vphantom{\dagger}}\vert n_{a}\rangle=\sum_{\bm{k}}\Bigl\{N_{\bm{k}}^{(a)\vphantom{\dagger}}+2\lvert\beta_{\bm{k}}\rvert^{2}\Bigl(N_{\bm{k}}^{(a)\vphantom{\dagger}}+\frac{1}{2}\Bigr)\Bigr\}\,.
\end{equation}
We clearly see the second term on the righthand side corresponds to the stimulated production of particles, in reference to \eqref{E:nvkdjfsd}.  In particular, the created particles obey the probability distribution
\begin{equation}
	P(n_{\bm{k}})=\frac{1}{\lvert\alpha_{\bm{k}}(\mathfrak{t}_{f})\rvert^{2}}\frac{\lvert\beta_{\bm{k}}(\mathfrak{t}_{f})\rvert^{2n_{\bm{k}}}}{\lvert\alpha_{\bm{k}}(\mathfrak{t}_{f})\rvert^{2n_{\bm{k}}}}
\end{equation}
after the parametric process, where $n_{\bm{k}}$ is the number of created particles in mode $\bm{k}$. Moreover, since $\beta_{\bm{k}}(t)$ can be expressed in terms of the fundamental solution of the equation of motion of the field, it  contains information of the field's evolution under the parametric process.

The fundamental solutions $d^{(i)}_{\bm{k}}(t)$ and their derivatives can be used to quantify the degree of nonadiabaticity $\mathfrak{N}_{\bm{k}}$ of each field mode
\begin{equation}\label{E:bgksfkfsd}
	\mathfrak{N}_{\bm{k}}(\mathfrak{t}_{f})=\frac{1}{2\omega_{i}\omega_{f}}\,\dot{d}^{(1)2}_{\bm{k}}(\mathfrak{t}_{f})+\frac{\omega_{f}}{2\omega_{i}}\,d^{(1)2}_{\bm{k}}(\mathfrak{t}_{f})+\frac{\omega_{i}}{2\omega_{f}}\,\dot{d}_{\bm{k}}^{(2)2}(\mathfrak{t}_{f})+\frac{\omega_{i}\omega_{f}}{2}\,d^{(2)2}_{\bm{k}}(\mathfrak{t}_{f})\,,
\end{equation}
where $\mathfrak{t}_{f}$ denotes the moment the parametric process ends. It is essentially the ratio of the energy of each field mode at the end of an arbitrary transition to the counterpart energy for the adiabatic process if the field mode is initially in its ground state.

We  now turn to the dynamics of the detectors in the parametric field.

\section{Dynamics of detector in an evolving  quantum field}

With these ingredients at hand, we are ready to discuss the nonequilibrium dynamics of an oscillator-detector coupled to  a quantum field which evolves in time.  {We use the simplest case of parametric frequency modulation to illustrate how the past history of evolution may register in  the detector's observables, and then give a brief discussion of more generic cases.}

In the detective scenario, suppose we have only one detector located at fixed $\bm{z}$ whose internal degree of freedom $\hat{Q}(t)$ has a constant oscillator frequency. We further assume that the coupling strength $e$ is a constant which does not change in time. Then the equation of motion \eqref{E:gbksddf} is simplified to
\begin{equation}\label{E:irinffgdd}
	\ddot{\hat{Q}}(t)+\Omega^{2}_{b}(t)\,\hat{Q}(t)-\frac{e^{2}}{M}\int_{0}^{t}\!ds\;G_{R,0}^{(\Phi)}(t-s;\bm{z})\,\hat{Q}(s)=\frac{e}{M}\,\hat{\Phi}_{h}(\bm{z},t)\,,
\end{equation}
with
\begin{equation}
	\frac{1}{2}\,\langle\bigl\{\hat{\Phi}_{h}(\bm{z},t),\hat{\Phi}_{h}(\bm{z},t')\bigr\}\rangle_{\textsc{st}}=G_{H,0}^{(\Phi)}(t,t';\bm{z})\,,
\end{equation}
We introduce a shorthand notation, $G_{H,0}^{(\Phi)}(t,t';\bm{z})=G_{H,0}^{(\Phi)}(\bm{z},t;\bm{z},t')$, when the two-point function is evaluated at the same spatial point $\bm{z}$. Here we have shifted the time origin such that $t=0$ is the moment when the coupling between the field and the internal degree of freedom of the detector (the harmonic oscillator) is switched on. According to earlier discussions, the field will behave like a squeezed field, after the time $t>\mathfrak{t}_{f}$ (now $\mathfrak{t}_{f}<0$), with a constant but mode-dependent squeeze parameter $\zeta_{\bm{k}}$. Moreover, for mathematical simplicity, we assume that the initial state of the total system takes a product form, and the initial state of the harmonic oscillator has the properties
\begin{align}
	\langle\hat{Q}(0)\rangle&=0\,,&\langle\hat{P}(0)\rangle&=0\,,&\frac{1}{2}\,\langle\bigl\{\hat{Q}(0),\hat{P}(0)\bigr\}\rangle&=0
\end{align}
where $\hat{P}$ is the canonical momentum conjugated to the displacement operator $\hat{Q}$, and $\langle\cdots\rangle$ is the expectation value taken with respect to the initial state of the detector's internal degree of freedom.

At the same spatial point, the retarded Green's function of the field in \eqref{E:irinffgdd} reduces to
\begin{equation}
	G_{R,0}^{(\Phi)}(\sigma;\bm{z})=-\frac{\theta(\sigma)}{4\pi}\Bigl\{2\delta'(\sigma)+\frac{m}{\sqrt{\sigma^{2}}}\,J_{1}(m\sqrt{\sigma^{2}})\Bigr\}\,,\label{E:turrnfjgdd}
\end{equation}
where now $\sigma=t-t'$, and $\theta(\sigma)\operatorname{sgn}(\sigma)=\theta(\sigma)$. The first term on the righthand side restricts the effect mediated by the field to the future lightcone of any source coupled with the field, and  since the source is massive it cannot travel at a speed as high as the lightspeed, hence the first term represents a local effect in the equation of motion \eqref{E:irinffgdd}. In fact, it will also contribute to a frequency renormalization. The second term has an additional effect not seen for a massless field. Since it is always nonzero inside the future lightcone of the source, it will induce a time-delayed effect on the source itself at later times. It implies that the effect is history-dependent and thus non-Markovian. These will be  most clearly seen when we write \eqref{E:irinffgdd} as
\begin{equation}\label{E:bdfhdd}
	\ddot{\hat{Q}}(t)+\Omega^{2}_{\textsc{r}}(t)\,\hat{Q}(t)+2\gamma\,\dot{Q}(t)+2\gamma\int_{0}^{t}\!ds\;\frac{\mathfrak{m}_{f}}{t-s}\,J_{1}(\mathfrak{m}_{f}(t-s))\,\hat{Q}(s)=\frac{e}{M}\,\hat{\Phi}_{h}(\bm{z},t)\,,
\end{equation}
for $t>0$, where we recall that $\gamma=e^{2}/(8\pi M)$, $M$ the mass of the oscillator's internal degree of freedom and $\mathfrak{m}_{f}$ the effective mass of the field due to the parametric process. We observe that when the mass $\mathfrak{m}_{f}$ of the field quantum goes to zero, the fourth term on the lefthand side vanishes. On the other hand, in general it does not vanish for $t>s$, that is, in a timelike interval, so the massive field can induce a non-Markovian effect on a detector, meaning,  the imprints of a detector on the field at earlier moments will affect the same detector at the present moment. Thus the evolution of the detector depends on its own past history. It is clearly seen here that most of the effects will come within the past interval of the order $\mathfrak{m}_{f}^{-1}$, with the effective strength proportional to $\gamma\mathfrak{m}^{2}_{f}$. We observe that the nonMarkovian effect of a more massive field in this configuration has a shorter range, but its effective strength increases faster with the effective mass $\mathfrak{m}_{f}$.

Similar to the field described in the previous section, we can construct a special set of homogeneous solution to \eqref{E:bdfhdd}, $d_{Q}^{(1)}(t)$ and $d_{Q}^{(2)}(t)$, which satisfy the initial conditions
\begin{align}
	d_{Q}^{(1)}(0)&=1\,,&\dot{d}_{Q}^{(1)}(0)&=0\,,&d_{Q}^{(2)}(0)&=0\,,&\dot{d}_{Q}^{(2)}(0)&=1\,.
\end{align}
When the physical frequency $\Omega_{\textsc{r}}$ is not a function of time, they take the rather simple forms
\begin{align}
	d_{Q}^{(1)}(t)&=e^{-\scriptstyle{\upsilon} t}\Bigl[\cos\varpi t+\frac{\scriptstyle{\upsilon}}{\varpi}\,\sin\varpi t\Bigr]\,,&d_{Q}^{(2)}(t)&=\frac{e^{-\scriptstyle{\upsilon} t}}{\varpi}\,\sin\varpi t\,,
\end{align}
with
\begin{align}
	\mathfrak{z}&=-\sqrt{\smash[b]{\gamma}^{2}-\Omega^{2}_{\gamma}\mp2\smash[b]{\gamma}\sqrt{\smash[b]{\mathfrak{m}_{f}^{2}}-\Omega^{2}_{\gamma}}}\,,&\upsilon&=+\operatorname{Re}\mathfrak{z}\,,&\varpi&=-\operatorname{Im}\mathfrak{z}\,, 
\end{align}
where $\Omega^{2}_{\gamma}=\Omega_{\textsc{r}}^{2}-\gamma^{2}$ is the resonance frequency of this driven system \eqref{E:bdfhdd}.  
If we make the Taylor expansion of $\mathfrak{z}$ with respect to the small $\mathfrak{m}_{f}$, we obtain
\begin{equation}
	\mathfrak{z}\simeq-\gamma\pm i\Omega_{\gamma}+\Bigl[\frac{\gamma}{2(\Omega^{2}_{\gamma}+\gamma^{2})}\mp i\,\frac{\gamma^{2}}{2\Omega_{\gamma}(\Omega^{2}_{\gamma}+\gamma^{2})}\Bigr]\,\mathfrak{m}_{f}^{2}\,.
\end{equation}
When $\mathfrak{m}_{f}/\Omega_{\gamma}$ is small, we see that $d_{Q}^{(1,2)}(t)$ are not much different from their counterparts of the detector coupled to a massless field. However, they are distinct in characteristics from their massless-field correspondences $d_{\bm{k}}^{(1,2)}(t)$ in two aspects. First, their amplitudes decay with increasing time due to dissipation, which is the reactive force of the radiation field \cite{QRad} emitted by the internal degree of freedom of the detector from its nonstationary motion. This will bring in another interesting feature we will cover later. Second, since the internal degree of freedom of the detector has a history-dependent dynamics, its motion at any moment will  depend on all of its earlier states accumulatively, as well as on the field configurations over the history. This has some profound implications if the detector is coupled to the  field before the onset of the parametric process of the field. The parametric process of the field will leave imprints on the nonequilibrium dynamics of $\hat{Q}(t)$.  In the full  duration before the detector's internal degrees of freedom reach complete equilibration, the time dependencies of various physical observables of the detector could allow us to decrypt {certain} information of the time dependence of the parametric processes of the field, via the nonlocal term and the noise force in the detector's dynamics.

Even if the detector is coupled to the field after the end of the field's parametric process, we can still extract partial information about the field. We have argued that at the end of the process, the field acquires mode-dependent but constant squeezing with respect to its initial configuration. The overall effect of squeezing can be read off from the relaxation dynamics of the detector, in particular, from the covariance matrix elements of the detector's internal degree of freedom such as  displacement or momentum dispersion. The finer information about the mode dependence of squeezed field can even be extracted if the detector responds only to a specific narrow band of the spectrum of the field.

Richer information can be extracted when we have multiple, well separated detectors. In addition to the fact that we can receive the information each detector reports about the field configuration at its location, the detectors can also exert mutual influence on one another \cite{RHA,CPR}.  As seen from \eqref{E:gbksddf}, we will have additional nonlocal influences from the other detectors. The motion of the internal degree of freedom of one detector will induce radiation of the field, which will propagate to the other detectors according to the nonlocal terms in the equation of motion \eqref{E:gbksddf}. All told, the dynamics of the detectors can be extremely complicated, with multiple scales and structures, {but  the challenge is exactly in this --  the retrieval of these embedded information} to reconstruct the field's parametric process. 

As an illustrative example of the {\it mutual nonMarkovian} effect, we consider a simpler configuration, where two uncoupled UD detectors are coupled to the same real  {\it massless} thermal scalar field of initial temperature $\beta^{-1}$~\cite{LinHu09,HHEnt,PRE18}. In this case, the self-nonMarkovian effect  in the massive system studied here is absent, so the effects of the mutual nonMarkovian influence stands out. In~\cite{HHEnt}, mutual nonMarkovianity introduces a fractional change of the order $\sin\Omega_{\gamma}\ell/(\Omega_{\gamma}\ell)$ to the damping constant, and of the order 
\begin{equation}
	\frac{\gamma}{\Omega_{\gamma}}\frac{\cos\Omega_{\gamma}\ell}{\Omega_{\gamma}\ell}\,,
\end{equation}
to the oscillating frequency of the normal modes of the internal degrees of freedom of the two detectors, when $\Omega_{\gamma}\ell\gg1$ and $\gamma/\Omega_{\gamma}<1$. Here $\ell$ is the distance between these two detectors. Although these corrections are usually small, their  decaying sinusoidal character in distance could be a distinguished signature for  identifying this mutual non-Markovian effect. Consider, for example,  how the mutual influence affects the detector's observables. We see that the total mechanical energy  becomes
\begin{equation}
	E(\infty)=\frac{1}{2}\operatorname{Im}\int_{-\infty}^{\infty}\!\frac{d\kappa}{2\pi}\;\coth\frac{\beta\kappa}{2}\biggl\{\frac{\Omega_{\gamma}^{2}+\kappa^{2}}{\Omega_{\gamma}^{2}-\kappa^{2}-i2\gamma\kappa-\dfrac{2\gamma}{\ell}\,e^{i\kappa\ell}}+\frac{\Omega_{\gamma}^{2}+\kappa^{2}}{\Omega_{\gamma}^{2}-\kappa^{2}-i2\gamma\kappa+\dfrac{2\gamma}{\ell}\,e^{i\kappa\ell}}\biggr\}
\end{equation}
once their motion is fully relaxed.  {The mutual influences, the $\ell$ dependence, appear in the denominators in the integrand.}


\section{Summary and discussions}

We now can summarize our findings and respond quantitatively to the questions raised at the beginning (see Abstract): a) Is it possible to access the nonMarkovian information contained in a Witness $W$ modeled by a UD detector  \textit{evolving with} a quantum field in an expanding universe? b) To what extent can a `Detective' $D$  detector \textit{introduced today} decipher the past history of the universe through the memories imprinted in the quantum field? We describe the scenarios below,  and,  to cover a broader scope,  include cases with two or more  detectors  in our discussions,  the analysis of which we hope to provide in  future communications. 

\subsection{Witness scenario}
In the Witness scenario, as the detectors evolve in progression with the parametric variation of the field, they record moment by moment all the accumulated effects the field have on their past histories. By tracking the time dependence of the detectors' physical observables one can gather a fair amount of knowledge about the parametric process of the field. One can assume different classes of time dependence of the scale factors: power law, as in radiation-dominated universe, exponential, as in inflationary universe and determine the functional form of the parametric process (the squeezing) of the quantum field.  We can formally express the two-point functions of the field in terms of the fundamental solutions $d_{\bm{k}}^{(1,2)}$ as in \eqref{E:bfkfkrte} and \eqref{E:brtyufb}. They in turn allow us to use \eqref{E:irinffgdd} to find the fundamental solutions $d_{Q}^{(1,2)}$ for the internal degrees of freedom, from which we can construct the detector responses or other physical observables of the detectors. These responses can be worked out as templates for different classes of cosmic expansions.  

\subsection{Detective scenario}

For the Detective scenario, the available information is more limited because the effects of the parametric process of the field is condensed into the complex squeezing parameter $\zeta_{\bm{k}}$, which is mode-dependent, but independent of the spacetime location (This can be taken as the leading order in a quasi-local expansion) after the process ends. Similar to the aforementioned scheme, the squeeze parameter can be related to $d_{\bm{k}}^{(1,2)}$ at the end of the parametric process, $t=\mathfrak{t}_{f}$ by \eqref{E:gbkrjbkdg1}, \eqref{E:gbkrjbkdg2} or \eqref{E:bgksfkfsd}, so it may allow us to fix the values $d_{\bm{k}}^{(1,2)}(\mathfrak{t}_{f})$. However they are not enough to determine the details of the parametric process except for some simple cases, but power law and exponential time dependence of the scale factor are simple enough cases used in practice.

As precautions we mention several  caveats in the  information retrieval capabilities of quantum detectors in these  nonMarkovian processes: i)  {We have only} finite time retrieval  {or finite resolution} of information in nonMarkovian processes, ii) The mapping from the functional form of the parametric process  {of the field} to the responses of the detectors may not be retro one-to-one  {due to (a) coarse-graining of the field dynamics and (b) the relaxation process in the detector}, so the reconstruction of the parametric process of the quantum field may be limited to general patterns, rather than sharp 1-1 correspondences.   iii) Since the detector is a quantum mechanical system the retrieval of the records in the detector without collapsing its state requires some skillful manipulations, but there are ways in quantum optics and quantum information designed for these purposes. With two  or more detectors one can obtain additional information of the quantum field such as the quantum correlation and entanglement properties.

\subsection{Severe limitations of perturbative treatment}

{We commented in the Introduction on the severe limitations of perturbative treatments of memories.   One is tempted to resort to perturbation methods if a small parameter like the coupling constant between the detector (system) and the field (bath) is present in the theory.  However,  an immediate concern is that such an approach will usually lose its precision as the evolution time prolongs because the coupling constant is related to the  relaxation time scale. The late-time result of the reduced system obtained by perturbative analysis is usually unreliable. In particular, take \eqref{E:bdfhdd} as an example, a low-order perturbative treatment easily misses the renormalization in the theory and the dissipative backreaction in the equation of motion. Both tend to appear in higher-order considerations than the fluctuation backreaction does. Thus the results obtained from low-order perturbative treatments  often show a characteristic growth in time. 

Physically, this is very similar to free diffusion. For the system described by \eqref{E:bdfhdd}, it is not difficult to see how this growth arises. Let us ignore the $\mathcal{O}(e^{2})$ term for the moment, then \eqref{E:bdfhdd} describes a undamped, noise-driven Brownian system. Our experience with the Brownian motion tells us that the displacement or the velocity dispersions usually grow with time\footnote{The validity of this statement actually depends on the bath dynamics and the form of the coupling between the system and the bath. It is true for the Ohmic bath of a scalar field we are considering. But for a supra-Ohmic bath, the growth may be more tamed~{\cite{HL11}}.}. After some time, these dispersions will be sufficiently large such that contributions from the $\mathcal{O}(e^{2})$ term in \eqref{E:bdfhdd} can no longer be neglected. A detailed analysis will show that it tends to counteract the effect of the fluctuation force from the bath, and may reduce the rate of further dispersive growth. For the linear system we are considering here this likely leads to equilibration. However, for a more complicated reduced system like a nonlinear oscillator, other things can happen. For example, the damping force may not be able to dissipate fast enough the energy stored in the nonlinear potential, which is presumably bounded from below but has been excited by the fluctuation force, for the system dynamics to dwindle down to the lowest part of the potential.}

\subsection{Related problems}

To broaden the scope of our investigation further, what we have studied here for cosmology can be applied to {\it dynamical Casimir effects} (DCE)~\cite{DCE} which share the same underlying theoretical structure. For DCE, a moving mirror squeezing the quantum field produces particle pairs. However, there are  differences in the key questions asked. In DCE the drive and its functionality are explicitly introduced in the design of the experiments, thus no one would be asking the kind of `archeology' question so important in cosmology: what means do we have to find out about the universe's past?  In DCE a more natural question one would  ask is,  how would a detector in a quantum field respond after it is squeezed by a moving mirror?  The nonMarkovian  quantum stochastic equations governing the detector's internal degrees of freedom  derived here can be of use for the analysis of the experimental data.  The other aspect which we have not discussed but had been addressed in cosmology involves the drive dynamics. If its trajectory is not a given fixed function of time but a dynamical variable, then it is even closer to the cosmological condition, where the scale factor describing the universe's expansion obeys the Einstein equations which needs to be solved self-consistently with the field equations. This is known as the cosmological backreaction problem \cite{HuVer20}. Interesting questions asked in cosmology have  exact analogies in DCE,  and answers provided in one can inspire the other.

To end this discussion we return to our detector. In our derivations we have {implemented} a stationarity condition for the UD detector after full relaxation. That implies the existence of a \textit{fluctuation-dissipation relation} in a quantum Brownian oscillator interacting with a squeezed quantum field. This relation is proven in a companion paper~\cite{FDRSq}, the existence of  which ensures this condition can be achieved in the detector. Such a relation for a quantum Brownian oscillator (or a harmonic atom) in a squeezed thermal field has many applications, as discussed  in~\cite{FDRSq}.  {The foundational and more complex issue of nonMarkovianity in cosmology, especially in the light of quantum wave functions or trajectories,  has yet to be more fully addressed.} \\\\

\noindent\textbf{Acknowledgment}  J.-T. Hsiang is supported by the Ministry of Science and Technology of Taiwan under Grant No.~MOST 110-2811-M-008-522.











\end{document}